\documentclass[twocolumn]{aastex63}

\shorttitle{ALMA formation height}
\shortauthors{Martinez-Sykora et al.}

\usepackage{natbib}
\newcommand{\jms}[1]{\color{black}{#1}}

\newcommand{\gol}{{\em gol\_lte}}
\newcommand{\golneq}{{\em gol\_nei}}
\newcommand{\mgk}{\ion{Mg}{2} 2796~\AA\ k3}
\newcommand{\longacknowledgment}{We gratefully acknowledge support by NASA grants NNX16AG90G, NNX17AD33G, 80NSSC18K1285 and contract NNG09FA40C (IRIS), NSF grant AST1714955. 
JdlCR is supported by grants from the Swedish Research Council (2015-03994), the Swedish National Space Board (128/15) and the Swedish Civil Contingencies Agency (MSB). This project has received funding from the European Research Council (ERC) under the European Union's Horizon 2020 research and innovation programme (SUNMAG, grant agreement 759548). 
The simulations have been run on clusters from the Notur project, and the Pleiades cluster through the computing project s1061, s1630, and s2053 from the High End Computing (HEC) division of NASA. This study has been discussed within the activities of team 399 ``Studying magnetic-field-regulated heating in the solar chromosphere" at the International Space Science Institute (ISSI) in Switzerland. To analyze the data we have used IDL. 
This research is also supported by the Research Council of Norway through 
its Centres of Excellence scheme, project number 262622, and through 
grants of computing time from the Programme for Supercomputing. Data are courtesy of IRIS, SDO/AIA and SDO/HMI. IRIS is a NASA small explorer mission developed and operated by LMSAL with mission operations executed at NASA Ames Research Center and major contributions to downlink communications funded by ESA and the Norwegian Space Centre. {\jms ALMA is a partnership of ESO (representing its member states), NSF (USA) and NINS (Japan), together with NRC (Canada) and NSC and ASIAA (Taiwan) and KASI (Republic of Korea), in cooperation with the Republic of Chile. The Joint ALMA Observatory is operated by ESO, AUI/NRAO and NAOJ. The National Radio Astronomy Observatory is a facility of the National Science Foundation operated under cooperative agreement by Associated Universities, Inc. We used the publicly available data ADS/JAO.ALMA\#2016.1.00202.S.}}

\begin{document}

\title{{\jms The Formation Height of Millimeter-wavelength Emission in the Solar Chromosphere}}

\correspondingauthor{Juan Martinez-Sykora}
\email{juanms@lmsal.com}

\author{Juan Mart\'inez-Sykora}
\affil{Lockheed Martin Solar \& Astrophysics Laboratory,
3251 Hanover St, Palo Alto, CA 94304, USA}
\affil{Bay Area Environmental Research Institute,
NASA Research Park, Moffett Field, CA 94035, USA.}
\affil{Rosseland Center for Solar Physics, University of Oslo, P.O. Box 1029 Blindern, N-0315 Oslo, Norway}
\affil{Institute of Theoretical Astrophysics, University of Oslo,
P.O. Box 1029 Blindern, N-0315 Oslo, Norway}

\author{Bart De Pontieu}
\affil{Lockheed Martin Solar \& Astrophysics Laboratory,
3251 Hanover St, Palo Alto, CA 94304, USA}
\affil{Rosseland Center for Solar Physics, University of Oslo, P.O. Box 1029 Blindern, N-0315 Oslo, Norway}
\affil{Institute of Theoretical Astrophysics, University of Oslo,
P.O. Box 1029 Blindern, N-0315 Oslo, Norway}

\author{Jaime de la Cruz Rodriguez}
\affil{Institute for Solar Physics, Department of Astronomy, Stockholm University, AlbaNova University Centre, SE-106 91, Stockholm, Sweden}

\author{Georgios Chintzoglou}
\affil{Lockheed Martin Solar \& Astrophysics Laboratory,
3251 Hanover St, Palo Alto, CA 94304, USA}
\affil{University Corporation for Atmospheric Research, Boulder, CO 80307-3000, USA}

\begin{abstract}
  In the past few years, the ALMA radio telescope has become available for solar
  observations. ALMA diagnostics of the solar atmosphere are of
  high interest because of the theoretically expected linear
  relationship between the brightness temperature at mm wavelengths and the local gas temperature in the solar atmosphere. Key
  for the interpretation of solar ALMA observations is understanding
  where in the solar atmosphere the ALMA emission
  originates. Recent theoretical studies have suggested that ALMA bands at 1.2 (band 6) and 3~mm (band 3)
  form in the middle and upper chromosphere at significantly
  different heights. We study the formation of ALMA diagnostics using a 2.5D radiative
  MHD model that includes the effects of ion-neutral interactions
  (ambipolar diffusion) and non-equilibrium ionization of hydrogen and
  helium. Our results suggest that in active regions and 
  network regions, observations at both wavelengths most often
  originate from {\jms similar} heights in the upper
  chromosphere, contrary to previous results. Non-equilibrium ionization increases the opacity
  in the chromosphere so that ALMA mostly observe spicules and fibrils
  along the canopy fields. We combine these modeling results with
  observations from {\jms IRIS, SDO and ALMA} to suggest a new
  interpretation for the recently reported ``dark chromospheric
  holes'', regions of very low temperatures in the chromosphere.
\end{abstract}


\section{Introduction} \label{sec:intro}

To better understand the origin of heating and dynamics in the solar chromosphere, it is important to reliably diagnose thermodynamic and magnetic field conditions in this important region in the solar atmosphere \citep[for a review][]{Carlsson:2019ARA&A..57..189C}.  Typically, observational constraints in the chromosphere are derived from spectral lines that are optically thick and formed under conditions of non Local Thermodynamic Equilibrium (non-LTE), such as \ion{Ca}{2} 8542\AA\  \citep{Cauzzi:2009km}, H$\alpha$ \citep{Rutten:2008ASPC..397...54R,Leenaarts:2012cr} or
\ion{Mg}{2} h 2803\AA\ and k 2796\AA\  \citep{Shimit:2015ApJ...811..127S}. The
interpretation of these diagnostics can be complicated, as the line formation
depends on complex radiative transfer effects such as partial frequency
redistribution (PRD) and 3D scattering
\citep[e.g.,][]{Leenaarts:2012cr}, as well as on time dependent
ionization \citep[at least for \ion{Ca}{2} and
H$\alpha$,][]{Wedemeyer-Bohm:2011oq,Leenaarts:2007sf,Leenaarts:2013ij}.
ALMA observations potentially offer an attractive alternative (or rather
complement, given the paucity of ALMA solar observations), as they
do not suffer from some of these effects. 

The advent of solar observations at ALMA has led to several recent
publications that summarize the potential of radio observations to
provide direct measurements of the plasma temperature for a wide range of
heights in the chromosphere \citep[see
][review]{Wedemeyer:2016SSRv..200....1W}. Such measurements would provide
novel diagnostics of chromospheric physical processes and direct
constraints on state-of-the-art numerical models of the chromosphere.
However, for a proper interpretation of ALMA observations, it is important to
understand where the diagnostics originate, especially given the
highly dynamic state of the chromosphere \citep[which is strongly
impacted by, e.g., magneto-acoustic shocks][]{Carlsson:1997ys}.

Current best estimates of
the formation height of ALMA diagnostics and the relationship between
observed brightness temperature and local gas temperature
\citep{Wedemeyer:2016SSRv..200....1W} are based on 3D radiative MHD (rMHD)
models of relatively quiet regions \citep{Carlsson:2016rt} in which
hydrogen is treated in non-equilibrium ionization (NEI)
\citep{Loukitcheva:2015A&A...575A..15L}. These models suggest
that there is a good relationship between brightness temperature and
local temperature, and that the various ALMA bands are formed at
different heights in the low to upper chromosphere. Such models have
also been used to analyze the benefits of combining ALMA and NUV observations from the
Interface Region Imaging Spectrograph \citep[IRIS,
see][]{De-Pontieu:2014yu} in order to derive semiempirical models from
the inversion of the observed intensities \citep{daSilvaSantos:2018A&A...620A.124D}. Similarly,
\citet{Loukitcheva:2015A&A...575A..15L} used a 3D rMHD simulation to assess the
potential of using ALMA observations to study chromospheric magnetic
fields. However, the numerical simulations utilized in those
publications were only representative of quiet Sun conditions. In addition,
these models
have not simultaneously included the effects of ion-neutral interactions in the
partially ionized chromosphere, time-dependent ionization and/or
missing physical processes such as the formation of type II spicules.

For the first time, we analyze the formation of ALMA intensities from
a very high spatial resolution simulation that is representative of
the dynamics, magnetic field configuration and fine structuring of
plage and strong network regions on the Sun. The simulations utilized in the present study
include time-dependent ionization of both hydrogen and
helium, interactions between neutral and ionized particles,
and the full stratification of the atmosphere from the upper
convection zone to the corona. To better understand the effects of
time dependent ionization, we use two simulations, one with and one
without NEI. We describe briefly the numerical
simulations (Section~\ref{sec:sim}) and ALMA synthetic calculations
(Section~\ref{sec:syn}). In Section~\ref{sec:res} we describe our
results and show that the formation height and the integration along
the line-of-sight (LOS) of the brightness temperature are highly dependent
on the electron density which is drastically increased in the upper
chromosphere as a result of time dependent ionization and the
increased mass loading resulting from spicular flows. These results
dramatically change the interpretation of ALMA observations,
indicating that in many regions these are dominated by fibrils and spicules
along the magnetic canopy, bringing them more in line with
expectations from theoretical approaches inspired by H$\alpha$
observations \citep{Rutten:2017A&A...598A..89R}. We also discuss how
our results offer a new interpretation for the recently discovered
``chromospheric holes'' \citep{Loukitcheva:2019ApJ...877L..26L}
and finish with conclusions (Section~\ref{sec:con}). 

\section{Numerical simulations}\label{sec:sim}

We use the two different 2.5D rMHD numerical simulations
analyzed in \citet{Martinez-Sykora:2019hhegol}. These simulations
have been calculated with the 3D rMHD {\it Bifrost} code
\citep{Gudiksen:2011qy} including scattering
\citep{Skartlien2000,Hayek:2010ac,Carlsson:2012uq}, thermal conduction
along the magnetic field, and ion-neutral effects, i.e., ambipolar
diffusion and Hall term
\citep{Martinez-Sykora:2012uq,Martinez-Sykora:2017gol,Nobrega-Siverio:2019tec}.
The simulations differ in their treatment of the ionization balance:
the \gol\ simulation is in LTE, while the \golneq\ simulation computes the ionization
balance in non-equilibrium for hydrogen and helium \citep{Leenaarts:2007sf,Golding:2014fk}.

In both simulations, the numerical domain covers a region that is
$90$~Mm wide and that covers a height
range from 3~Mm below to 40 Mm above the photosphere. The horizontal
resolution is uniform with a $14$~km grid spacing, while the vertical
resolution is non-uniform with the largest resolution in the
photosphere, chromosphere and transition region ($\sim 12$~km grid
spacing). The magnetic field configuration includes two plage regions
of opposite polarity with an unsigned mean magnetic field of
$\sim190$~G, and loops connecting both polarities (Figure~\ref{fig:tau}A). 

The boundary conditions are periodic in the horizontal direction and
open in the vertical direction, allowing waves and plasma to go
through. In addition, the bottom boundary has a constant entropy in
regions of inflow to maintain the solar convective motions with
$\sim 5780$~K effective temperature at the photosphere. Further
details on the setup and analysis of these two simulations can be
found in \citet{Martinez-Sykora:2019hhegol} and
in \citet{Martinez-Sykora:2017gol} for the \gol\ simulation. 

\section{Synthesis of ALMA observations}\label{sec:syn}

To compute synthetic observations from our simulations in the ALMA observations at 1.2 (ALMA band 6) and 3~mm (ALMA band 3),
we used the LTE module in the {\jms Stockholm inversion code (STiC)} code
\citep{de-la-Cruz-Rodriguez:2016tg,delaCruzRodriguez:2019A&A...623A..74D}. STiC utilizes
the electron densities and gas pressure stratifications from the
simulations to compute the partial densities of all species that are
involved in the calculations. Continuum
opacities are calculated using routines ported from the ATLAS code
\citep{Kurucz:1970SAOSR.309.....K}, which include the main opacity source
at mm wavelengths (free-free hydrogen absorption, see
\citealt{Wedemeyer:2016SSRv..200....1W}). The emergent intensity
is calculated using a formal solver of the unpolarized radiative
transfer equation based on cubic-Bezier splines
\citep{Auer:2003ASPC..288....3A,delaCruzRodriguez:2013ApJ...764...33D}.  

\section{Results}\label{sec:res}

\subsection{Formation height of ALMA observations}\label{sec:where}

\begin{figure*}
	\includegraphics[width=0.95\textwidth]{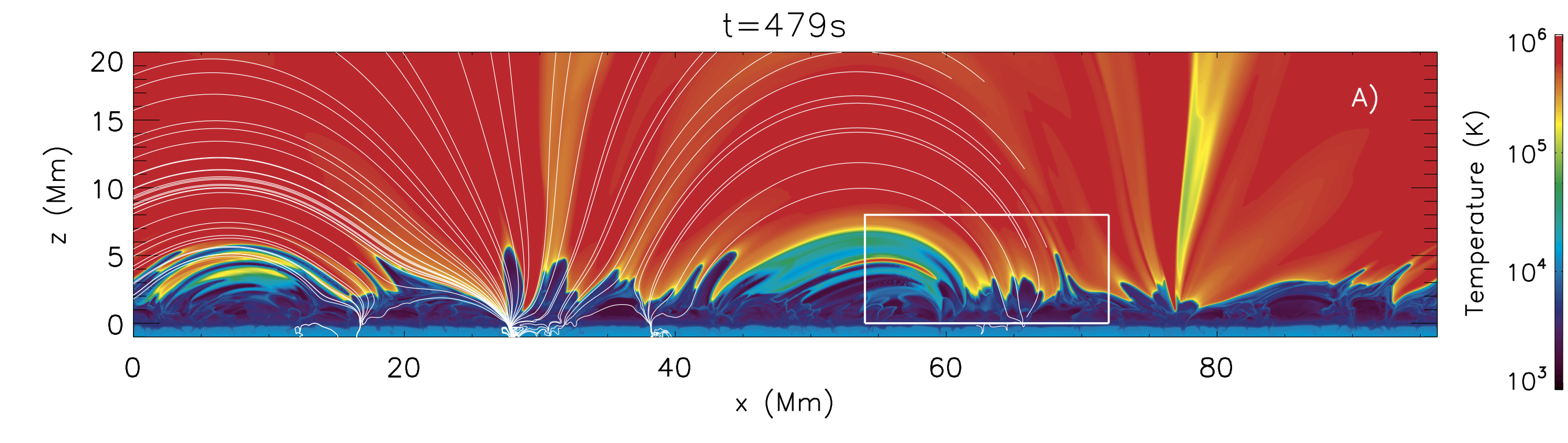}  
	\includegraphics[width=0.95\textwidth]{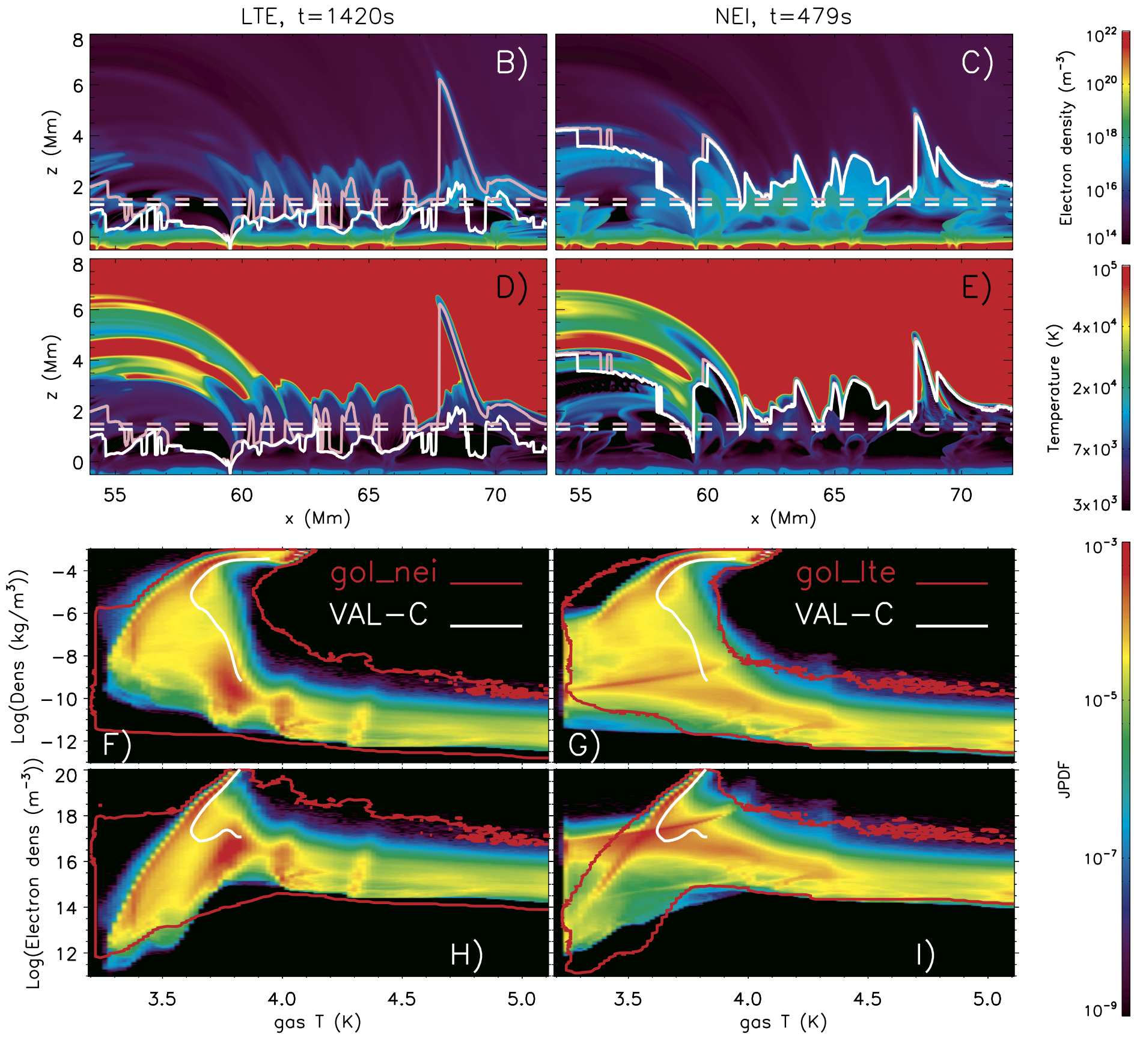}  
	\caption{\label{fig:tau} The formation height of ALMA
          passbands depends on the thermodynamic state of the
          chromosphere which is impacted significantly by ion-neutral
          interactions and NEI. Panel A
          shows the temperature in the \golneq\  simulation.
          For clarity, magnetic field lines are shown only in the left half of the
          numerical domain. Panels B-E zoom in on the white box in
          panel A, with maps of electron density (B, C) and
          temperature (D, E) for the \gol\ (left), and \golneq\
          (right) simulations. Formation height ($\tau=1$) of  observations at 3
          and 1.2mm are shown with pink and white lines, respectively. The
          dashed lines correspond to $\tau=1$ for the VAL-C model for
          the corresponding mm observations. Panels F-I show joint
          probability density functions (JPDF) of temperature and
          density (F, G) and of temperature and electron
          number density (H, I), each computed from a time series of 12
          minutes of solar time. For comparison between both
          simulations, we include the red contours which correspond to the
          temperature and density regime (of the whole simulation) at
          JPDF$=5\times 10^{-5}$ for the other simulation (see
          labels). The white line corresponds to the VAL-C model.}
\end{figure*}

In order to address the typical formation height of ALMA observations,
we computed the optical depth ($\tau$) at 1.2 and 3mm wavelengths
(ALMA bands 6 and 3, respectively). Figure~\ref{fig:tau} shows maps of
electron number density (panels B and C) and temperature (panels D and
E) for the \gol\ (left), and \golneq\ (right) simulations. Overplotted
are white and pink solid lines which show the heights for which
$\tau=1$ for observations at 3 and 1.2mm, respectively. Under LTE conditions,
the $\tau=1$ heights at 3mm are well separated from $\tau=1$
heights at 1.2mm. The latter typically occur within $0.5-2$~Mm above the photosphere, i.e., lower-mid chromosphere and often form within
the cold expanding bubbles produced in the wake of magneto-acoustic
shocks. {\jms Averaged over 12 minutes in the numerical simulation, the average formation height at wavelengths of 1.2mm is 0.9~Mm (with a standard deviation of 0.7~Mm).} Observations at 3mm are formed at significantly greater heights and often
forms along type II spicules. {\jms As a result, the average formation height at 3mm is 1.8~Mm (standard deviation of 1~Mm).  The mean difference of the formation heights at these wavelengths is 0.92 ($\pm 0.85$~Mm).}

Under NEI conditions, the electron density (panels C and I) is much higher
within the chromosphere than in LTE (panels B and H). {\jms On average the 1.2mm emission is formed at a height of $2.67\pm1.08$~Mm, while the 3mm emission is formed at a mean height of $2.78\pm1.09$~Mm. The difference in average formation heights of these wavelengths is thus $0.11\pm0.3$~Mm.}This can be explained
as follows. In NEI, the recombination timescales are much longer than
the MHD timescales. This means that during the passage of shocks (a key
constituent of chromospheric dynamics), the cooling from expanding
bubbles in the wake of shocks leads to a decrease of the plasma
temperature instead of the recombination that would occur under LTE
conditions. Consequently, the formation height for both wavelengths is moved to significantly greater heights in the upper-chromosphere, near the
transition region. In fact, most of the time and almost
everywhere these two wavelengths observe very similar regions: low-lying
loops, fibrils or/and spicules. The impact of NEI on the formation
height is thus significant and fundamentally alters the interpretation
of the ALMA observations.

For comparison we add, for both ALMA wavelengths, the height at which the optical depth is unity
for a VAL-C \citep{Vernazza:1981yq} atmosphere (dashed horizontal
lines in Figure~\ref{fig:tau}B-E). It is clear that the large
variability of the formation height of these two wavelengths in a rMHD model is not captured by the VAL-C model.

It is important to note that the chromosphere is highly structured,
with large temperature and density (or electron density) variations.
This is clearly shown with the Joint Probability Distributions
Functions (JPDF) of temperature and density, and temperature and
electron number density shown in Figure~\ref{fig:tau}F-I. Note that
the JDPF's axes are in logarithmic scale.  We refer the
reader to \citet{Martinez-Sykora:2019hhegol} for details on the
differences of these thermal properties between the \gol\ and \golneq\ simulations.

The temperature variations within the chromosphere are greatest in the
\golneq\ simulation because any heating or cooling due to various
entropy sources (e.g., ambipolar heating or work) change the
temperature instead of recombining or ionizing the plasma.
The VAL-C model cannot reveal these variations (white line in
Figure~\ref{fig:tau}F-I). The three preferred temperatures at
$\log(T/K) = 3.8$, 4 and 4.3 in the LTE case (panels F-G) correspond
to the ionization temperatures of hydrogen and helium
\citep{Leenaarts:2007sf,Golding:2016wq}. In NEI, these three bands
smear out (panels H-I). In addition, in NEI, plasma seems to follow
an adiabatic relation ($T=[10^{3.2},10^4]$~K,
$\rho=[10^{-10},10^{-8}]$~kg~m$^{-3}$, and
$n_{e}=[10^{17},10^{18}]$~m$^{-3}$). This is due to fact that the
cooling from the expansion in the wake of acoustic shocks (in the low
chromosphere, along spicules, and along low-lying loops) will not lead
to recombination (because of the long timescales involved in NEI).
As mentioned above, this leads to a much larger electron density and opacities in NEI than in LTE
(up to 4 orders of magnitude). So, the NEI changes completely the
electron density distribution within the chromosphere and therefor
the formation height at 3 and 1.2mm as shown in panels B-E.

\subsection{Relationship between ALMA brightness temperature and plasma temperature}\label{sec:what}

\begin{figure*}
    \includegraphics[width=0.95\textwidth]{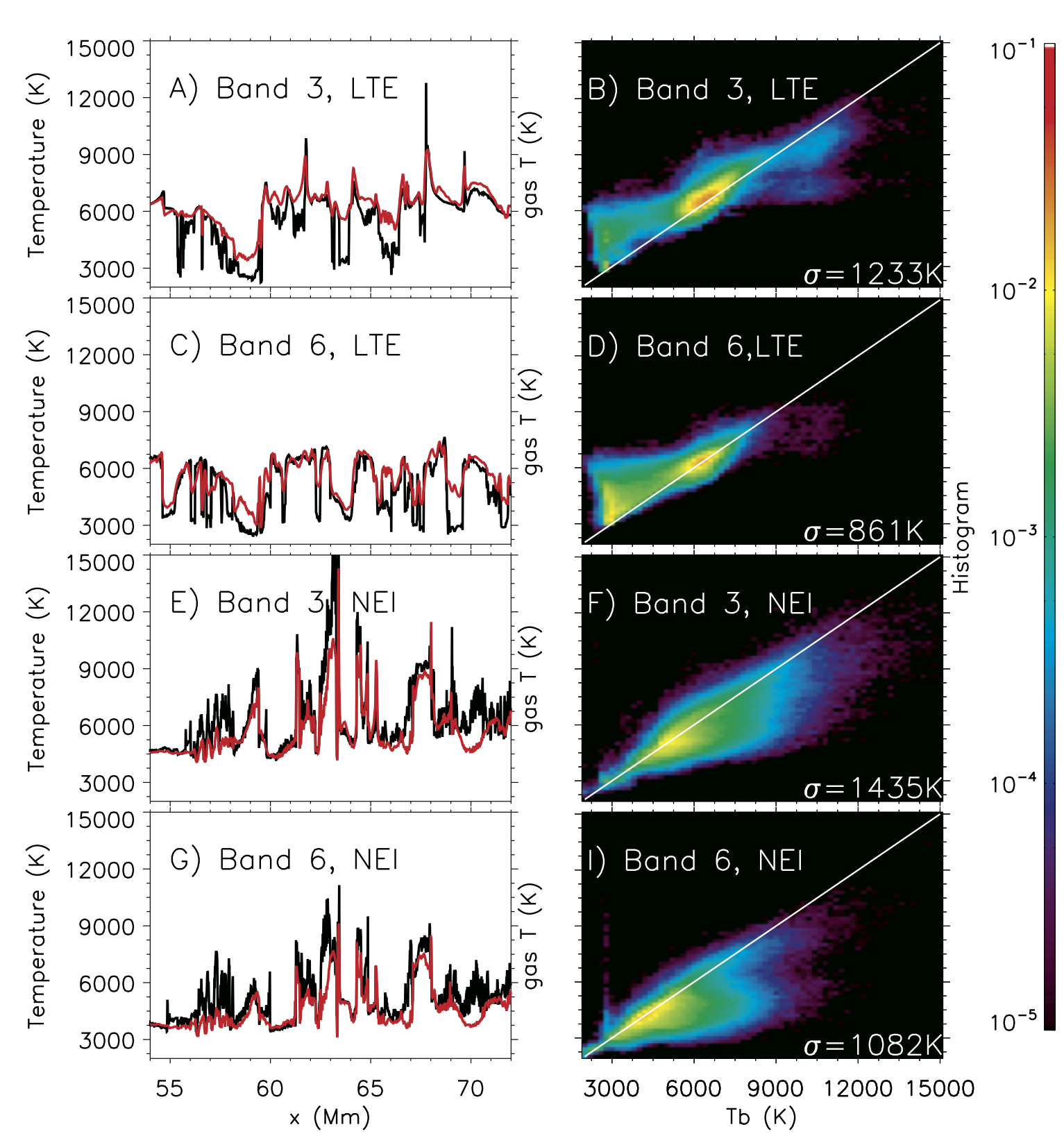}
	\caption{\label{fig:syn} Synthetic ALMA observations (red) can differ from the gas temperature at $\tau = 1$ (black) by several $10^3$~K. Synthetic ALMA observations at 3mm ({\jms odd} rows) and at 1.2mm (even rows) are shown for for \gol\ (top two rows) and \golneq\ (bottom two rows). Left column shows the JPDF between the  synthetic ALMA observations (x-axis) and gas temperature at $\tau = 1$ (y-axis).}
\end{figure*}

Given that NEI changes the formation height of ALMA observations, we
now consider the diagnostic capability of ALMA in our models. In
particular, we want to address whether the observed brightness
temperature ($Tb$) is correlated with the local gas temperature at
$\tau=1$. The left column of Figure~\ref{fig:syn} shows the synthetic
$Tb$ at 3 and 1.2mm (red) and the gas temperature at their
corresponding $\tau=1$ (black). Observations at 3mm show greater variability in space than at 1.2mm, both in LTE and NEI. One can see that there is
some correlation between the two temperatures. However, in several
locations the discrepancy between the two temperatures, in both LTE
and NEI, can reach up to a few $10^3$~K. 

To further illustrate this, we calculated the JPDF of the gas temperature at $\tau=1$
and $Tb$ (right column in Figure~\ref{fig:syn}). The JPDFs show some
correlation between the two temperatures, which visually appears to be
somewhat better for the NEI case.
However, the standard deviation of the difference between $Tb$ and
the gas temperature ($\sigma$, bottom-right labels in the right column
of Figure~\ref{fig:syn}) is larger in NEI than in LTE. Observations at 1.2mm provide
a better match with the gas temperature than at 3mm. Still, the
correlation is far from perfect and limits the degree to which ALMA
observations can constrain numerical models. The discrepancy between
the two temperatures is caused by the LOS integration
as detailed below. 

\begin{figure*}
    \includegraphics[width=0.99\textwidth]{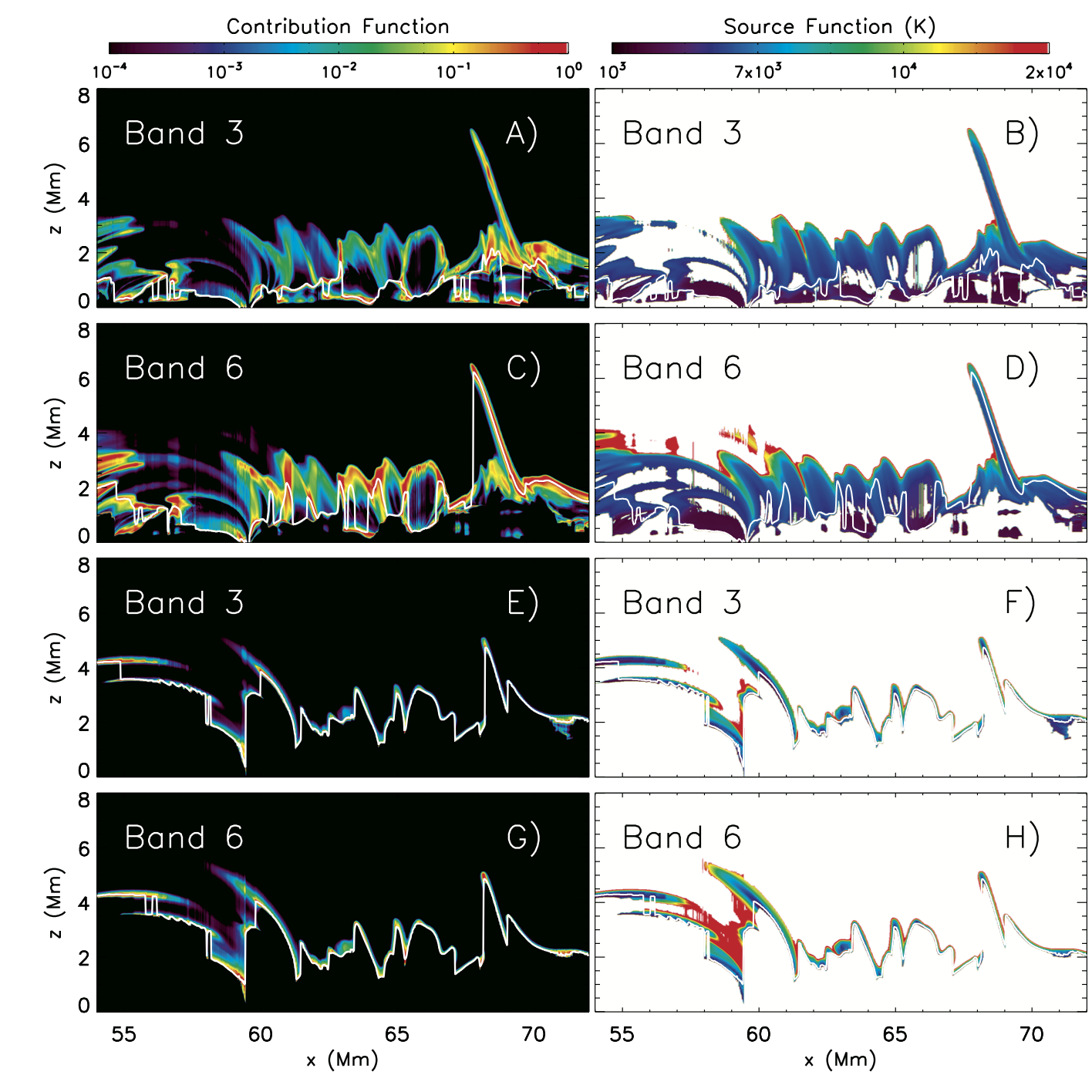}
	\caption{\label{fig:sf3lte} The impact from LOS integration on
          the synthetic observations at 1.2 (even rows) and 3mm
          ({\jms odd} rows) for the \gol\ (two top rows) and \golneq\
          simulations (two bottom rows). Left column shows the
          contribution function normalized to the highest values along
          the LOS ($CF$). Right column is the source function where
          we masked regions with $CF<10^{-4}$. The formation height
          ($\tau=1$) of the corresponding observations are shown with white lines.}
\end{figure*}

To investigate the LOS effects, we calculated the contribution
function and the source function for both wavelengths
(Figure~\ref{fig:sf3lte}). In LTE, the contributions to the total
intensity are formed over a wide range of heights, often with parcels
at very different heights equally contributing. As we know, the
ionization in LTE is highly underestimated because, in
NEI, the recombination time-scales are larger than the timescales
related to magneto-acoustic evolution or associated with other entropy
sources \citep[e.g., ambipolar
heating,][]{Martinez-Sykora:2019hhegol}.
The fact that in LTE very different packets of plasma along the LOS
contribute to the brightness temperature leads to the lower degree of
correlation between these two temperatures (as compared to the NEI case).

In contrast, the contribution function for the NEI case is more confined to a narrow
region along the LOS: $\tau=1$ occurs at much greater heights, which
significantly reduces the number of plasma elements above that height
that can contribute. Despite the general visual impression of a
somewhat better overall correlation between brightness and plasma temperature for the NEI case, we
nevertheless find a larger standard deviation $\sigma$ (i.e., worse
correlation). This is because in the \golneq\ simulation extremely
sharp and large variations arise in the source functions in
comparison to the \gol\ simulation (right column of
Figure~\ref{fig:sf3lte}). These are caused by the stronger temperature
gradients within the chromosphere in the \golneq\ simulation
\citep[see][for details]{Martinez-Sykora:2019hhegol}. These in turn
are caused by the fact that any heating or cooling changes the gas
temperature instead of ionizing or recombining the plasma. This
results in large temperature variations rather than preferentially
keeping the plasma around the ionization temperatures (Figure~\ref{fig:tau}F-I).  

In summary, since the formation height of the ALMA observations
occurs at greater heights in NEI than in LTE, the LOS superposition
is much smaller for the former. However, this is counteracted by the
fact that the \golneq\ simulation has sharper transitions in
temperature. As a result, even if the LOS is integrated over a narrower
region, the LOS effects become more important.

\begin{figure}
    \includegraphics[width=0.49\textwidth]{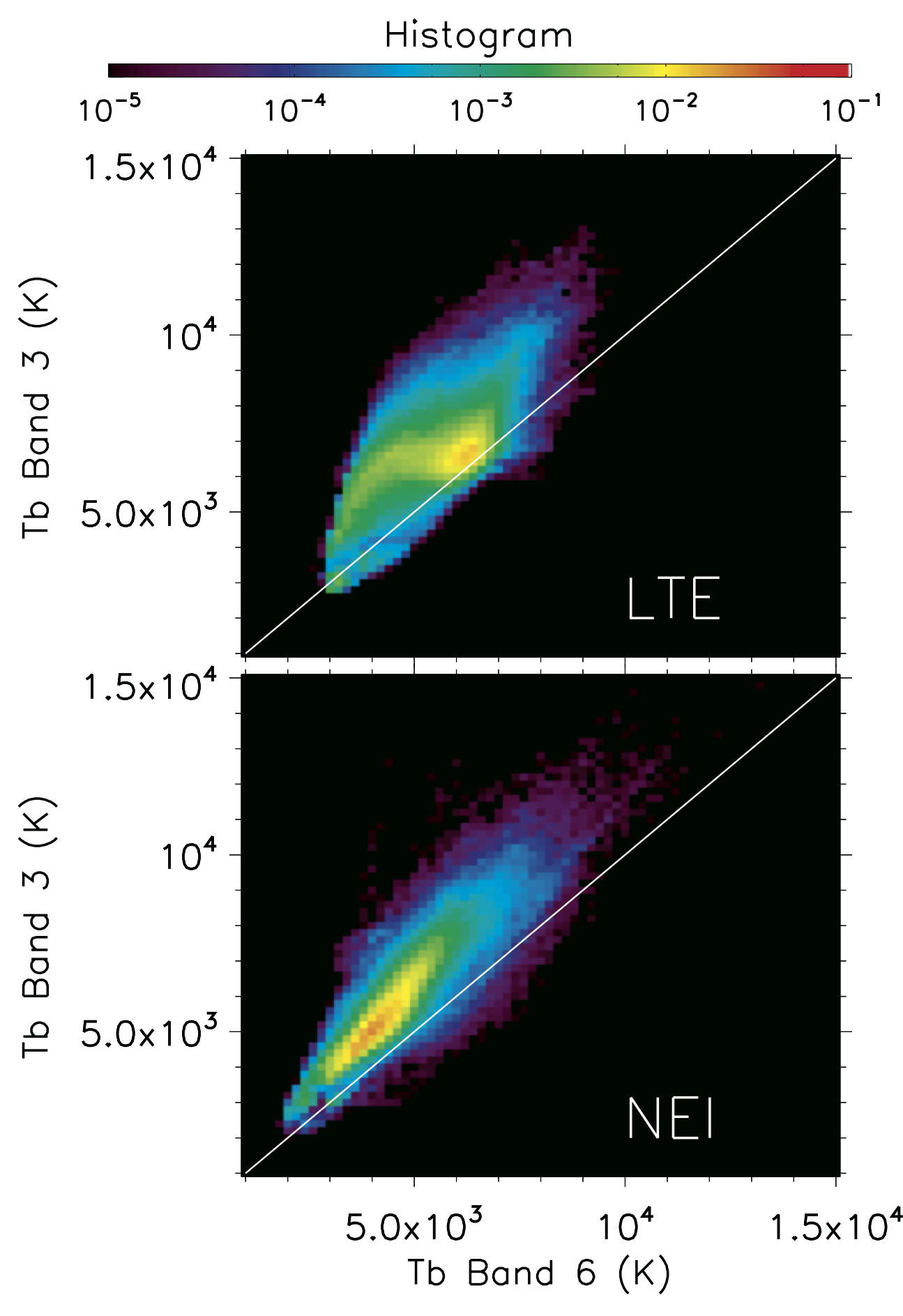}
	\caption{\label{fig:b36jpdf} The JPDF between the brightness
          temperatures at 3 and 1.2mm show a good correlation
          for the NEI case (bottom), in contrast to the LTE case (top).}
\end{figure}

Given the close proximity of the typical formation heights, the
question arises whether these different wavelengths have significantly
different diagnostic capability for the NEI case. Panels F and H in
Figure~\ref{fig:sf3lte} show that the source function is very similar
for both wavelengths. It is then not surprising that the JPDF between at 3 and 1.2mm (Figure~\ref{fig:b36jpdf}) shows a strong correlation for
the \golneq\ simulation (contrary to the LTE case). The lack of
correlation in the LTE case is expected (top panel), as these wavelengths are formed in
very different regions. However, in NEI, the correlation between 3 and 1.2mm is actually better than the correlation between $Tb$ and
gas temperatures shown in Figure~\ref{fig:syn}F and I. We find a
significant difference in average temperatures between 3 and 1.2mm, as the former is formed at slightly greater heights, essentially in
the same structures. The mean brightness temperature difference between the two wavelengths is $1280$~K. This results from strong temperature gradients within the structures (e.g, perpendicular to the magnetic field direction in
low-lying loops or inclined spicules). If these findings are borne out
by comparisons between 3 and 1.2mm observations (hampered by the
lack of simultaneity between wavelengths), our findings suggest that
observations at 1.2mm might be preferred, given the higher spatial
resolution that can be obtained using ALMA and the slightly better
correlation with gas temperature at $\tau=1$ (Figure~\ref{fig:syn}).
If 3 and 1.2mm observations were close to simultaneously 
possible, they might help identify locations of sharp temperature
gradients.

\subsection{Alternative observational interpretations}\label{sec:obs}

Our results can also be used to provide a new interpretation of recent
ALMA observations by \citet{Loukitcheva:2019ApJ...877L..26L} who reported
regions of low brightness temperature and named these ``chromospheric
holes'', suggesting a possible link to previous observations of
low-lying cool gas deduced from molecular CO lines.

\begin{figure*}
    \includegraphics[width=0.90\textwidth]{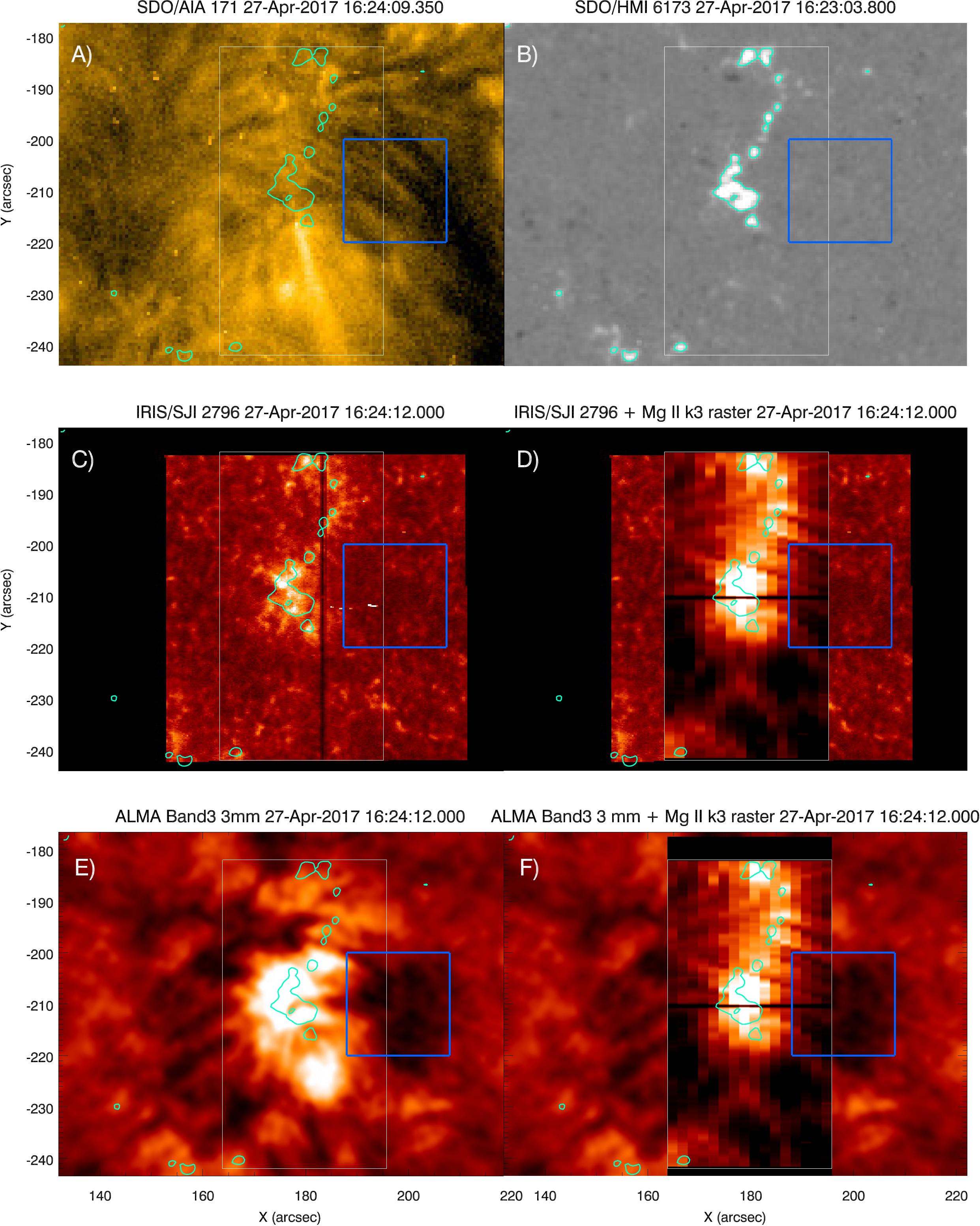}
	\caption{\label{fig:iris} Analysis of {\jms IRIS spectroheliograms, SDO/AIA-HMI and ALMA} observations suggests
          that low-lying fibrils occur in the same region as where
          \citet{Loukitcheva:2019ApJ...877L..26L} (in their Figure 1) reported the
          presence of a ``chromospheric hole'' region of low brightness
          temperature using ALMA observations. Panel A shows an
          SDO/AIA 171~\AA\ image; panel B shows an SDO/HMI
          magnetogram (scaled to $\pm 200$~G); panel C shows an IRIS 2796
          slit-jaw image, while panel D shows
          a spectroheliogram \jms{(averaged over the same 45 minute time period as the ALMA observations)} at the core of the \ion{Mg}{2} k line (superimposed on top of
          the IRIS 2796 slit-jaw image){\jms ; panel E shows the time-averaged relative brightness temperature in ALMA band3, while panel F includes
          the time-integrated \mgk\ spectroheliogram}. Green contours correspond to
          $100$~G in panel B and the blue box outlines the region of
          low ALMA brightness temperature from \citet{Loukitcheva:2019ApJ...877L..26L} (see their Figure~1).}
\end{figure*}

Here we present a different possible scenario using our simulations
and combining with observations using SDO/AIA \citep{Lemen:2012uq}{\jms,
IRIS and ALMA band3 observations}. Both ALMA and IRIS observed the same region through a
coordinated ALMA/IRIS campaign (Figure~\ref{fig:iris}). The IRIS
observing program was centered at heliocentric coordinates of
170\arcsec, -210\arcsec, with a medium (i.e., 60\arcsec\ FOV along the
slit), coarse (i.e., 2\arcsec\ steps), 16-step
raster with an exposure time per slit position of 2s and a raster
cadence of 32s. {\jms The ALMA interferometric data were acquired on 2017/04/27 in Band 3 (at 3mm, i.e., 100 GHz) in configuration C40-3 \citep[see][for further details]{Loukitcheva:2019ApJ...877L..26L}. ALMA obtained 10.5 minute scans separated by 2min calibration scans, with 2s integration time, for a total of 37min within 16:00-16:45 UT (45 minutes). The ALMA data we show here is averaged over that time interval. ALMA solar observations are detailed in \citet{Shimojo:2017SoPh..292...87S,White:2017SoPh..292...88W}}.

While it is difficult to determine the morphology of the chromospheric hole region
from the 2796 SJI images, the IRIS spectroheliogram at the core
of the \mgk\ line shows clear evidence of long fibrils, commonly
seen outlining low-lying canopy fields originating from stronger field
regions. The SDO/HMI magnetograms (panel B and green contours) confirm
that these fibrils do indeed connect to a strong magnetic field region
with significant magnetic field strength ($>100~$G), i.e., a decayed
plage or enhanced network region. In addition, timeseries of SDO/AIA 171\AA\ images
similarly reveal a mix of dark and bright features compatible with
low-lying fibrils in the ``chromospheric hole'' region (blue box). {\jms Detailed inspection of ALMA band3, the integrated-in-time IRIS \mgk\ spectroheliogram, and SDO/AIA 171 observations shows fibril-like features with similar morphology in all three observations. Further details of the ALMA observational analysis can be found in \citet{Loukitcheva:2019ApJ...877L..26L}.}

Given this observational context, our simulations indicate that such
areas of ``low-lying loops outlining the canopy that originates from
strong field regions'' should have low brightness temperatures. For
example, the region between 55 and 60 Mm (in x) in Figure~\ref{fig:tau} shows
that the low-lying loops are sites of very high electron density
(i.e., high opacity in ALMA) with
low temperatures, as low as 3,500-4,500K (Figure~\ref{fig:syn}). This is
very similar to what is reported for ``chromospheric holes'' by
\citet{Loukitcheva:2019ApJ...877L..26L}. In our simulations, the ALMA
observations of low temperatures arise from low-temperature
sub-threads in low-lying loops, a natural consequence of the
chromospheric dynamics when taking into account mass loading from
spicules, heating from shocks and ambipolar diffusion, and
NEI effects. For details on these structures, we
refer the reader to \citet{Martinez-Sykora:2019hhegol}. 

We also note that the high ALMA $Tb$ near the footpoints of the
fibrils \citep[Figure~1A in][]{Loukitcheva:2019ApJ...877L..26L}
matches the shape of the bright region in the \mgk\ spectroheliogram
(panel D). Our simulations suggest that these high temperatures may
be caused by the spicules and associated heating at the footpoints of
the low-lying fibrils.

\section{Conclusions and discussion}\label{sec:con}

We have used two different state-of-the-art 2.5D rMHD
simulations (one assuming LTE ionization, one assuming NEI), both including ion-neutral interaction effects, to
investigate the formation height and diagnostic capability of the ALMA
bands 3 (3mm) and 6 (1.2mm) observations. Our results show that NEI, and the strong mass loading in the upper chromosphere (arising
from heating caused by ambipolar diffusion, as well as spicules and
shocks), both have a significant impact on the interpretation of ALMA
observations.

In our NEI model, the formation height of ALMA at both wavelengths
occurs at greater heights than in LTE models. In addition, both wavelengths observe roughly the same features and region.

Previous studies \citep[e.g.,][]{Loukitcheva2:015ASPC..499..349L}
focused on understanding ALMA observations using
numerical models that did not include spicular mass loading or
ambipolar diffusion/heating \citep{Carlsson:2016rt}. They also
typically did not compare LTE versus NEI with the exception
of \citet{Leenaarts:2006ASPC..354..306L}. However, our results
seem to be contrary to \citet{Leenaarts:2006ASPC..354..306L}: it
is unclear in their results if the ALMA formation height in NEI is
at greater heights than in LTE. This is most likely because their
model is much shallower (only up to the lower chromosphere) and did
not include the greater densities in the upper chromosphere seen
in our 2.5D rMHD models that include ambipolar diffusion. As
a result, their model did not show the large electron density and opacities at greater heights
that appears in NEI, and it did not couple the NEI effects to
the hydrodynamics. 

Our 2.5D rMHD model including NEI differs from previous work
\citep[e.g.,][]{Wedemeyer-Bohm:2007vn,Loukitcheva2:015ASPC..499..349L}
in several ways: 1) our model includes more physical processes, i.e.,
non-equilibrium hydrogen and helium ionization, as well as ambipolar
diffusion; 2) our magnetic field configuration mimics a plage or
strong network region, while previous models typically mimicked very
quiet Sun and/or smaller numerical domains; 3) our model has
higher densities and opacities in the upper atmosphere due to the presence of
low-lying loops and spicules, 4) our model has higher spatial
resolution, e.g, more than four times better resolution than the
simulation used in \citet{Loukitcheva2:015ASPC..499..349L}.

Our models show that due to the high opacities in the upper
chromosphere from the presence of type II spicules, low lying loops
and large-scale magnetic field configuration, and taking into
account the NEI effects, the formation height of both  wavelengths is
located in the upper chromosphere.
Due to the hydrogen and helium NEI and the ambipolar diffusion, the plasma has very large
temperature variations along the LOS. 

Our results are well aligned with predictions from
\citet{Rutten:2017A&A...598A..89R} who theorizes that ALMA mm
observations will show opacities that are similar or larger than
H$\alpha$. Consequently, he predicts that ALMA will mostly observe
fibrils along the canopy, while anything below will be masked by these
fibrils. \citet{Molnar:2019ApJ...881...99M}  found that ALMA band 3 correlates nicely with H$\alpha$ core width. Our \golneq\ simulation which includes NEI, also
shows large ALMA opacities in the upper chromosphere which mask
anything below. Consequently, 
contributions at 1.2 and 3mm are confined to the upper chromosphere, i.e.,
low-lying loops (canopy fibrils), and spicules, instead of the
acoustic shocks in the lower atmosphere \citep{Wedemeyer:2016SSRv..200....1W}. 

Although the contribution  functions for both wavelengths are confined to a very narrow region
in the \golneq\ simulation, our NEI results show that care must be taken
when interpreting ALMA brightness temperatures as a local gas
temperature. Not only is there a large spread in the correlation
between these quantities, the plasma also shows very large temperature gradients along
the LOS since any cooling or heating will change the plasma
temperature instead of being amortized by ionization or
recombination. In addition, spicules and low-lying loops may contain
thin threads of very different
temperatures \citep{Martinez-Sykora:2019hhegol}. Our results suggest that  emission at 1.2mm and 3mm is formed at similar (but not identical) heights in the solar atmosphere. Consequently, the comparison between emission at these two wavelengths can provide information about temperature gradients (e.g., within the same feature). Our NEI model of active region and enhanced network shows a mean brightness temperature difference of 1280~K between the two wavelengths. This is similar to the difference in mean brightness temperature between the two ALMA bands of 1400 K in the {\jms averaged over the whole Sun} \citet{White:2017SoPh..292...88W}. {\jms Note that these observations include both AR and QS, although the latter dominates in terms of areal coverage.}

One main reason for the different results in the current work
(compared to previous work) is the fact that our simulations show large opacities 
in the upper chromosphere. Analysis has shown that this is caused by
several factors: the inclusion of ambipolar diffusion, as well as the
inclusion of both large-scale and small-scale magnetic field
structures. In previous work, due to the small numerical domain,
typically, the magnetic field expands drastically with height,
diluting shocks and other drivers of mass flows, so that
it has been very difficult to reach high densities in the upper
chromosphere \citep{Martinez-Sykora:2013ys,Carlsson:2016rt}. 

As with any numerical model, care should be taken when applying it to
the real Sun. We note that our model is limited to two dimensions, and
it is crucial to expand this model into 3D. Nevertheless, we expect that in plage and
strong network regions the field will not suffer as much expansion
with height as quiet Sun models in 3D. One should also keep in mind
that models tend to simplify the magnetic structure and may limit the
LOS superposition compared to what happens on the Sun.

Nevertheless and in conclusion, our results indicate that state-of-the-art inversions
and/or synthetic observations from rMHD models need to
take into account NEI effects for a proper interpretation
of ALMA observations.

\acknowledgements{\longacknowledgment} 

\bibliographystyle{aasjournal}
\bibliography{collectionbib}

\end{document}